\documentclass[a4paper, oneside, twocolumn, notitlepage, 10pt]{extarticle_ecoc}
\usepackage{ecoc}
\usepackage{subcaption}
\begin{document}
\selectlanguage{english}    


\title{Distributed Sensing on a 110-kV Overhead-Line Maintenance Operation on an Operational Optical Ground Wire}%


\author{
    Konstantinos Alexoudis\textsuperscript{(1,2)}, Torm Järvelill\textsuperscript{(3)}, Hendrik Johann Kerm\textsuperscript{(3)}, Kaida Kaeval\textsuperscript{(3)}, Florian Azendorf\textsuperscript{(1)}, \\ 
    Vincent Sleiffer\textsuperscript{(1)}, Jasper Müller\textsuperscript{(1)}, 
    Chigo Okonkwo\textsuperscript{(2)}, Tom Bradley\textsuperscript{(2)}
}

\maketitle                  


\begin{strip}
    \begin{author_descr}

        \textsuperscript{(1)} Adtran, Frauenhoferstraße 9a, 82152 Planegg, Germany, \textcolor{blue}{\uline{konstantinos.alexoudis@adtran.com}}
        
        \textsuperscript{(2)} High-Capacity Optical Transmission Laboratory, Eindhoven University of Technology, Netherlands

        \textsuperscript{(3)} Tallinn University of Technology, Estonia

    \end{author_descr}
\end{strip}

\renewcommand\footnotemark{}
\renewcommand\footnoterule{}


\begin{strip}
    \begin{ecoc_abstract}
        We demonstrate dual‑modal distributed sensing on an operational 110‑kV OPGW during crane maintenance. DAS resolves meter‑scale lifting events and matches impulsive events to phone audio, while DTSS quantifies post‑reclamping residual strain up to $\sim$398~$\mu\epsilon$, enabling maintenance verification and asset monitoring in transmission grids. ©2026 The Author(s) 
    \end{ecoc_abstract}
\end{strip}
\vspace{0.2\baselineskip}

\section{Introduction}
Optical ground wires (OPGW), combining a lightning shield conductor with embedded telecommunication fibers serving critical communications, are widely deployed across high-voltage transmission grids and are increasingly repurposed as distributed sensors \cite{Lu:2019}. While fibers in OPGW are increasingly used for environmental monitoring \cite{Carvalho:2019}, no method currently exists to remotely verify the mechanical outcome of maintenance operations on overhead lines. Grid operators rely solely on crew reports to confirm that work was completed correctly. Distributed acoustic sensing (DAS) based on phase-sensitive optical time-domain reflectometry ($\phi$-OTDR) can convert these fibers into dense vibration sensor arrays spanning tens of kilometers, without requiring dedicated sensor hardware at each tower or span \cite{Zhang:2025}. Complementary to DAS, Distributed temperature and strain sensing (DTSS) is used for assessing the quasi-static strain along the same cable, capturing slow mechanical changes that fall outside the acoustic bandwidth.

\begin{figure*}[b!]
    \vspace{-0.5\baselineskip}
    \centering    \includegraphics[width=\linewidth]{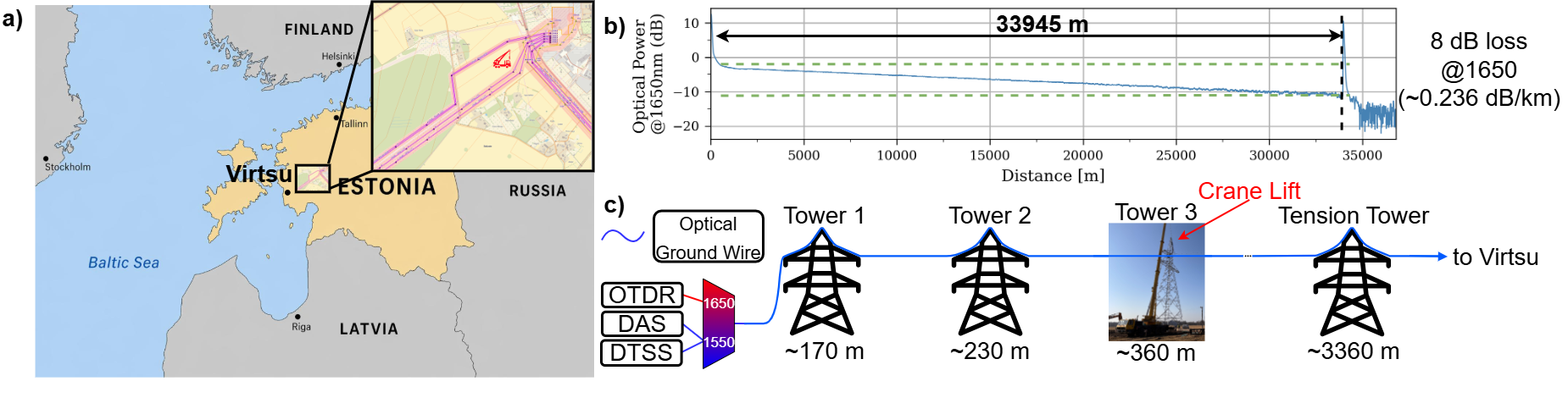}
    \caption{(a) Lihula–Virtsu 110-kV transmission corridor (overview) with zoom on the monitored OPGW section near Lihula substation. (b) OTDR trace at 1650~nm (100~ns pulses) showing $\sim$8~dB loss over the 33.9~km link. (c) Setup schematic; crane location at $\sim$360~m and tension tower at 3360~m.}
    \label{fig:setup}
\end{figure*}

To date, $\phi$-OTDR on OPGW has primarily been applied to passive environmental monitoring. In \cite{Ding:2021}, vibration reconstruction and ice accretion detection via OPGW is shown, while in \cite{Canudo_Overhead:2024} span-by-span sag extraction on a 110-kV transmission corridor is achieved and later classified mechanical intrusions above 20 Hz through simultaneous OPGW and optical phase conductor cable monitoring \cite{Canudo_Simultaneous:2024}. In parallel, $\phi$-OTDR a field trial on operational telecom fibers within power utility networks has demonstrated intrusion detection and electrical cable health monitoring via 50 Hz power-frequency acoustic signatures \cite{Rai:2025}, while electrical discharge event detection was shown in a lab environment \cite{Alexoudis:2025}.

However, all reported DAS studies, whether on OPGW or adjacent telecom fibers, focus on the detection of passive environmental events such as wind, ice loading, or faults. The acoustic and strain signatures of planned grid maintenance operations, which are routine yet mechanically significant events, have not been characterized with distributed fiber sensing. To the best of our knowledge, no published work has characterized the DAS and DTSS signature of a controlled mechanical operation on an operational overhead line. In this work, we report combined DAS and DTSS measurements on the Estonian 110-kV transmission grid, capturing a three-hour cable lifting operation. DAS resolves the full mechanical sequence, ground contact, crane lift, and re-attachment, and DTSS reveals a localized residual strain at the re-clamped tower, demonstrating that OPGW-based distributed sensing can capture the overhead line maintenance operations.

\section{Experimental Setup}
The measurements were made on the 110-kV high-voltage overhead transmission line between the Lihula and Virtsu substations in western Estonia (Fig.~\ref{fig:setup}a), using the embedded SMF in the OPGW. An OTDR trace using a 1650~nm laser with 100~ns pulse width confirms the link length of 33.9~km and a link loss of $\sim$8~dB (Fig.~\ref{fig:setup}b), without any visible events on the trace. The interrogator at the Lihula substation provides both DAS and DTSS, using a narrow-linewidth laser ($< \text{1 kHz}$) at 1550.12~nm \cite{Alexoudis:2026}, and is coupled with the OTDR using a 1550/1650~nm filter on the same fiber. The monitored section starts at the substation and passes a sequence of suspension towers, with the maintenance tower at $\sim$360~m and a tension tower at 3360~m where the OPGW is anchored with a dead-end clamp (Fig.~\ref{fig:setup}c).

For DAS operation, the system was configured to send pulses with a pulse width of 100~ns, corresponding to a nominal spatial resolution of 10~m. A gauge length of 10~m was selected for signal processing, providing fine-grained localization of strain transients along the 33.9~km link. The backscattered Rayleigh signal is received by a coherent receiver and sampled at 250~MS/s, resulting in approximately 0.4~m sample spacing. The interrogator measured continuously throughout the three-hour maintenance event with a pulse repetition frequency of 800 Hz. A 15–350~Hz bandpass filter is applied to the DAS data.

For DTSS operation, the same interrogator launches probe pulses into the fiber and captures the Brillouin backscattered light as a function of time, from which the Brillouin frequency shift (BFS) is extracted. A DTSS trace was acquired after the maintenance work was completed, with a pulse width of 100~ns, corresponding to a spatial resolution of approximately 10~m. The frequency of the local oscillator was swept within a range of 10.6~GHz to 11.0~GHz with a frequency resolution of 2~MHz and 3000 averages, to extract the BFS profile. In the absence of a pre-maintenance reference trace, spatial BFS variations relative to the neighboring undisturbed spans are interpreted as localized mechanical strain, noting that differential temperature contributions between spans cannot be fully excluded.  

On 10 March 2026, planned maintenance was performed at the $\sim$360~m tower (Fig.~\ref{fig:setup}a). The line was de-energized, isolated, and grounded, and all conductors, including the OPGW, were lowered to the ground. The DAS measurement started while the cables rested on the ground, capturing the crane lift, re-attachment, and post-maintenance data.

\section{Results}
\begin{figure}[h!]
    \centering
    \includegraphics[width=\columnwidth]{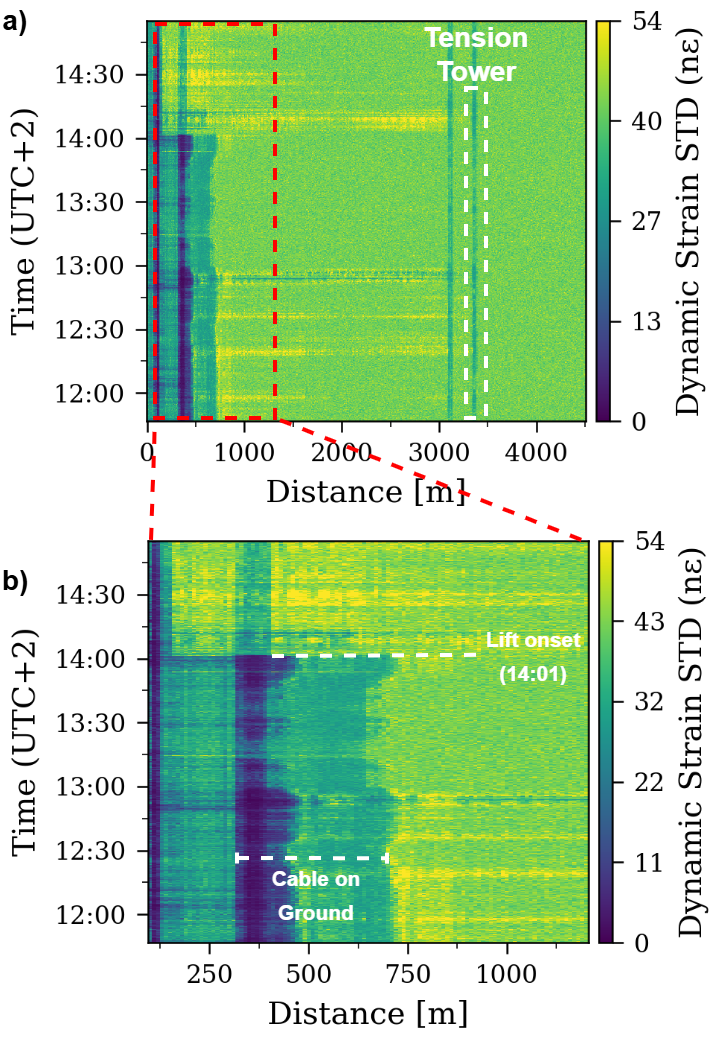}
     \caption{DAS dynamic strain std (1~s windows) along the Lihula–Virtsu OPGW. (a) Full 4.5~km overview; dashed line marks the tension tower at 3360~m. (b) Zoom on 100–1200~m. The baseline strain levels (320–740~m, before 14:01) indicate the ground-resting span; after the lift, only 320–390~m remains quiet.}
    \label{fig:das_heatmap}
\end{figure}

\noindent The top panel of Fig. \ref{fig:das_heatmap}a shows the dynamic strain along the first 4.5 km of fiber over the entire measurement period. The dynamic strain is 22-44~$n\epsilon$ in the aerial section, attributed to ambient mechanical excitation of the suspended cable. In the first 800 m of the link, the standard deviation of the phase (std) is lower, as illustrated in Fig.~\ref{fig:das_heatmap}b, where the maintenance happened. In this section, before approximately 14:01  (all times UTC+2), a broad corridor between 320 and 740 m shows a much lower phase variation than the surrounding fiber, with a phase std corresponding to a dynamic strain of $\sim$3~$n\epsilon$. After lift, the dynamic strain increases to 22~$n\epsilon$. This is the maintained span where the cables, including the OPGW, sat on the ground, shielded from the wind that dominates the acoustic signature of the overhead spans on either side.

Starting around 14:01, the horizontal features extend from $\sim$390~m to $\sim$3360~m, where they end abruptly in a tension tower. Unlike suspension towers, which allow longitudinal strain transients to propagate freely along the OPGW, tension towers anchor the cable with a dead-end clamp, blocking further propagation. A similar reduction in dynamic strain is observed at $\sim$3100~m, coinciding with another tower location. The underlying coupling mechanism remains subject to further investigation.

Fig.~\ref{fig:das_heatmap}b illustrates a zoomed view of the section from 100~m to 1200~m. Following the onset of the lift at 14:01, the low-variance zone contracts from the entire range of contact with the ground (320–740~m) to a residual section of $\sim$320–390~m that remains at $\sim$3~$n\epsilon$. This section, directly adjacent to the maintenance tower at $\sim$360~m, is mechanically constrained by the crane during the lift and therefore shows less sensitivity. The rest of the former quiet zone (390–740~m) rises to $\sim$22–44~$n\epsilon$, matching the level of vibration driven by the wind of the adjacent aerial spans, which is consistent with the fact that this portion of the span is freely suspended between the crane and the next tower.

\begin{figure}[h!]
    \vspace{-0.8\baselineskip}
    \centering
    \includegraphics[width=\columnwidth]{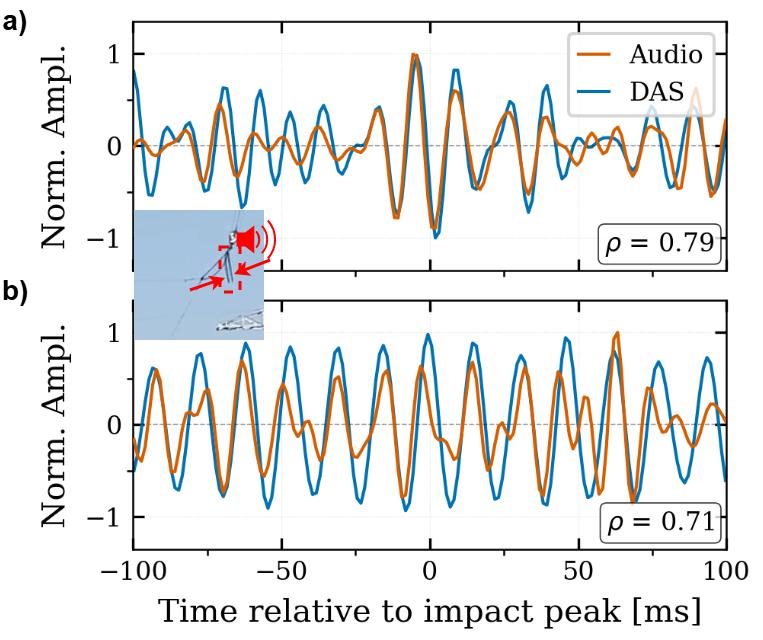}
    \caption{Normalized overlay of DAS (blue) and on-site audio (red) after 60--120~Hz bandpass filtering for two independent chain collision events: (a) 14:04:38, $\rho$=0.79; (b) 14:04:48, $\rho$=0.71.}
    \label{fig:audio_das}
\end{figure}

Additional on-site video footage captured the two ends of a steel chain, used to guide the cable, colliding with each other during the lift. To validate that the DAS transients reflect genuine mechanical events, we reconstructed an acoustic signal from the DAS phase at 360~m and matched it with two independent collision events to the on-site audio track by cross-correlation (Fig.~\ref{fig:audio_das}). Both signals were bandpass-filtered (60–120~Hz) and peak-aligned. The waveforms exhibit a sharp onset followed by oscillatory decay at a comparable dominant frequency of $\sim$65~Hz, with $\rho$ = 0.79 and 0.71, well above the level expected for two uncorrelated signals, suggesting that DAS phase reconstruction preserves sufficient acoustic fidelity to identify and distinguish mechanical events. We want to underline that clear audibility in the DAS playback confirms the matching of events as well, demonstrating the utility of DAS playback as a validation and potential classification tool.

\begin{figure}[h!]
    \vspace{-1.8\baselineskip}
    \centering
    \includegraphics[width=\columnwidth]{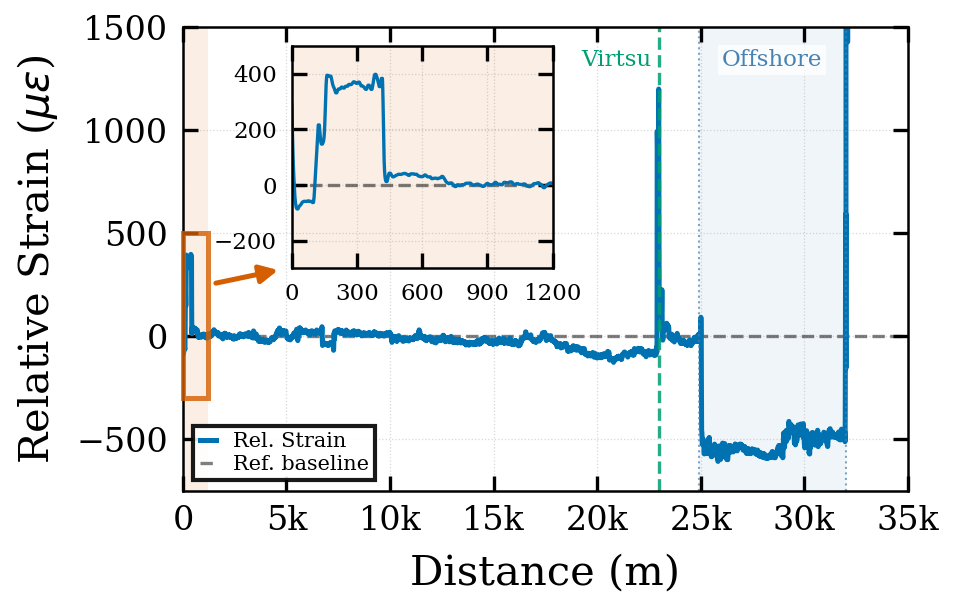}
    \caption{Relative strain versus distance. Similar strain, the 0~$\mu\epsilon$-line, is visible over the complete aerial link up to Virtsu location at $\sim$23~km. Afterwards, the fiber goes offshore (minus relative strain). The inset shows the first 1.2~km of OPGW, with a localized increase of up to $\sim$398~$\mu\epsilon$ in the 100 to 436~m section, indicating residual tensile strain from the re-clamping.}
    \label{fig:DTSS}
\end{figure}

Fig.~\ref{fig:DTSS} shows the relative strain vs. distance extracted from the BFS profile measured after completion of the maintenance work. The OPGW section up to Virtsu (at $\sim$23~km) is clearly visible, showing a close to 0~$\mu\epsilon$ value except for the section between 100 to 436~m (inset). Note that the offshore section, following after Virtsu, up to the end of the link, is also clearly visible, showing a lower BFS value, which is a combination of lower strain and likely temperature in this link section.
Since the OPGW section was measured simultaneously under identical thermal conditions, any deviation of the BFS from this baseline reflects mechanical strain rather than temperature. The negative relative strain below $\sim$100~m reflects the transition from the in-house launch fiber, which carries negligible mechanical pre-strain as a buried section, compared to the aerial OPGW spans. A localized strain increase is observed between approximately 100 and 436~m, which reaches its peak at $\sim$398~$\mu\epsilon$, which was derived using the standard Brillouin strain coefficient of 0.05~MHz/$\mu\epsilon$ for SMF \cite{Galindez}. The affected region extends beyond the maintained span towards the Lihula substation. This is consistent with the mechanical behavior of suspension towers, which allows longitudinal tension to be redistributed between adjacent spans.

\section{Conclusion}
We demonstrated the first DAS and DTSS measurement of a controlled cable lifting operation on a 110-kV high-voltage transmission line. The DAS captured dynamic strain variations from $\sim$3~$n\epsilon$ when the cable was on the ground, to $\sim$22-44~$n\epsilon$ when the cable was freely suspended between the spans. Chain collisions captured with the phone and DAS were effectively matched, showing potential for using DAS playback as a classification tool. Finally, DTSS revealed a localized residual strain change at the re-clamped tower of up to $\sim$398~$\mu\epsilon$ over a $\sim$336~m section, confirming an alteration of the static tension distribution. This shows that OPGW can serve not only as a communications backbone but also as a real-time asset monitoring tool for grid operators when using DAS and DTSS.

\clearpage
\section{Acknowledgements}
We acknowledge partial support from the Dutch Ministry of Economic Affairs and Climate Policy (EZK) through the PhotonDelta National Growth Fund Programme on Photonics and the Quantum Delta NL National Growth Fund Programme on Quantum Technology. This work was also partially funded by the European Commission under grant agreements 10113933 (ECO-eNet) and 101189703 (ICON). We further acknowledge the bilateral project "DistraSignalSense" between Eindhoven University of Technology, The Netherlands, and Adtran Networks SE. Finally, we thank the Estonian transmission system operator Elering AS for providing access to the 110 kV infrastructure and for their support during the field measurements.

\defbibnote{myprenote}{%
}\printbibliography[prenote=myprenote]

\vspace{-4mm}

\end{document}